\documentclass[%
 aip,
 amsmath,amssymb,
 reprint,%
]{revtex4-1}

\usepackage{graphicx}
\usepackage{dcolumn}
\usepackage{bm}

\usepackage[utf8]{inputenc}
\usepackage[T1]{fontenc}
\usepackage{mathptmx}
\usepackage{etoolbox}

\usepackage[english]{babel}     
\usepackage[table,xcdraw]{xcolor}
\usepackage{xcolor}
\usepackage{longtable}
\usepackage{graphicx}
\usepackage{amssymb}

\makeatletter
\def\@email#1#2{%
 \endgroup
 \patchcmd{\titleblock@produce}
  {\frontmatter@RRAPformat}
  {\frontmatter@RRAPformat{\produce@RRAP{*#1\href{mailto:#2}{#2}}}\frontmatter@RRAPformat}
  {}{}
}%
\makeatother
\begin{document}

\preprint{AIP/123-QED}

\title[The performance of HRPA(D) coupling constants]{On the performance of HRPA(D) for NMR spin-spin coupling constants: \\ Smaller molecules, aromatic and fluoroaromatic compounds}
\author{Louise M{\o}ller Jessen}

\author{Stephan P. A. Sauer}
\affiliation{%
Department of Chemistry, University of Copenhagen, DK-2100 Copenhagen Ø, Denmark.
}%

\date{\today}

\begin{abstract}
In this study, the performance of the doubles-corrected higher random-phase approximation (HRPA(D)) has been investigated in calculations of NMR spin-spin coupling constants (SSCCs) for 58 molecules with the experimental values used as the reference values. 
HRPA(D) is an approximation to the second-order polarization propagator approximation (SOPPA), and is therefore computationally less expensive than SOPPA.
HRPA(D) performs comparable and sometimes even better than SOPPA, and therefore when calculating SSCCs it should be considered as an alternative to SOPPA. 
Furthermore, it was investigated whether a CCSD(T) or MP2 geometry optimization was optimal for a SOPPA and a HRPA(D) SSCCs calculation for 8 smaller molecules. CCSD(T) is the optimal geometry optimization for the SOPPA calculation, and MP2 was optimal for the HRPA(D) SSCC calculations. \\

\end{abstract}

\maketitle

\section{Introduction}

For the calculation of molecular properties, like NMR spin-spin coupling constants (SSCCs), several methods are available such as CC3 \cite{Auer2001_CC3, AUER2003_CC3, AUER2009_CC3, Faber2017_SSCC_CC3_CCSDT}, CCSD \cite{Sekino1986_CCSD, Perera1994_CCSD, Perera1996_CCSD, NOOIJEN1997_CCSD, Perera2010_CCSD}, and SOPPA \cite{jod42_SOPPA-SSCC,jod130_SOPPA, spas030_SOPPA-SSCC, spas037, Schnack-Petersen_SOPPA_Approximation} that in general produce results in good agreement with experimental values.
However, these methods are computationally costly for larger molecules, and therefore cheaper alternatives are wanted.

The polarization propagator methods, doubles-corrected random-phase approximation (RPA(D)) and the doubles-corrected higher random-phase approximation (HRPA(D)),  \cite{spas025, Schnack-Petersen_RPAD_HRPAD} are both approximations of the second-order polarization propagator approximation (SOPPA). 
Both RPA(D) and HRPA(D) lower the computational cost but at the same time slightly lower the accuracy compared to SOPPA. 
HRPA(D), in which the most computationally expensive terms from the \textbf{B$^{(2)}$} matrix have to be evaluated in every iteration, only lowers the cost with around 10\%, however, the accuracy is comparable to SOPPA.
RPA(D) on the other hand lowers the computationally cost by around 75\%, however, the method has been found to be prone to have problems with triplet instabilities, which affect triplet properties such as the spin-spin coupling constant.\cite{Schnack-Petersen_RPAD_HRPAD, Moller2020_RPAD-HRPAD_SSCC_cycloalkanes} 

But how well do RPA(D) and HPRA(D) perform in the calculation of spin-spin coupling constants?
So far it has not been tested extensively for SSCCs in contrast to polarizabilities.\cite{spas192, spas201, spas203} 
The methods were implemented for SSCCs in 2018 by Schnack-Petersen et al. \cite{Schnack-Petersen_RPAD_HRPAD} and their performance was tested on 20 smaller mostly inorganic molecules in comparison with CCSD results. In their work, SOPPA performs better than HRPA(D), which performs better than RPA(D). But compared with HRPA and RPA both methods perform significantly better.
RPA(D) and HRPA(D) were then further studied in 2020 by Møller et al. \cite{Moller2020_RPAD-HRPAD_SSCC_cycloalkanes} for carbon-carbon SSCCs on 39 saturated cycloalkanes. 
The RPA(D) and HRPA(D) SSCCs were compared with the results of SOPPA SSCC calculations, where HRPA(D) performed better than RPA(D). 
In the article, the computational cost was checked compared to SOPPA. For these molecules, the cost of the HRPA(D) SSCC calculations was reduced between 15 and 65 percent, and for RPA(D) the cost of running the SSCC calculations was reduced between 60 to 85 percent.
In a recent study, \cite{Jessen_SSCC_solvent_effect_HRPAD-SOPPA} the HRPA(D) method was combined with the polarizable continuum model for treating the effect of solvents and SSCCs calculations in solution were performed with the HRPA(D) and SOPPA models for 20 molecules. 
In general, the HRPA(D) SSCCs were closer to the experimental values than the SOPPA SSCC in both the vacuum and the solvent calculation. 

In this study, it has now been investigated how well the HRPA(D) and SOPPA methods perform for the calculation of SSCCs of 58 molecules. 
The 58 molecules consist of two diverse groups: "smaller molecules" (31 molecules), and "aromatic and fluoro-aromatic molecules" (27 molecules). 
Some of the smaller molecules are taken from the set of molecules for which Faber et al.\cite{Faber2017_SSCC_CC3_CCSDT} had calculated SSCCs at the CCSD and even CC3 level of theory at geometries optimized at the  CCSD(T) level of theory. 
In the present study, SSCCs were then calculated with SOPPA and HRPA(D) at these CCSD(T) optimized geometries, which are then compared to the CCSD and CC3 results by Faber et al. \cite{Faber2017_SSCC_CC3_CCSDT}
Since the SSCCs for 7 of these molecules (CO, HCCH, FCCH, FCCF, F$_2$CO, H$_2$CO, and HCN) had been calculated for two different optimized geometries (CCSD(T) and MP2),
it is furthermore possible to discuss, which geometry optimization leads to better results in the calculation of SSCCs.

\section{Computational details}
The spin-spin coupling constants have been divided into three different groups. 
Set I and set II both consist of smaller molecules, where the geometries have been calculated with two different methods, CCSD(T) and MP2, respectively.
Set III consists of aromatic and fluoro-aromatic compounds. 

RPA(D) calculations have been carried out for all the molecules and the results can be found in supplementary materials. 
However, due to frequent triplet instabilities the results will not be commented on further in this paper. 
Only a few of the molecules did not show triplet instabilities, mostly consisting of organic molecules containing only single bonds. 

Even though adding solvation to the calculation affects the SSCCs,\cite{Ruud_Solvent_effect_SSCC_DFT-PCM, Mogelhoj_prediction_of_SSCC, Jessen_SSCC_solvent_effect_HRPAD-SOPPA} a solvation model was not used in the present study, as the solvation effects were previously found to almost equal in SOPPA and HRPA(D) calculations and thus cancel out in a comparison between the two methods.
Similarly, vibrational contributions have not been included in this study, even though the molecular vibrations affect the SSCCs, which several authors have investigated.\cite{Solomon1977_vib-avg, ruden2003vibrational, ruden2004nmr, nmr04-RudenRuud, Dracinsky2009_vib-avg, Sneskov2012_vib-avg, Faber2015_vib-avg, spasB12, Gleeson2023}

\textbf{Set I} contains the 7 molecules CO, HCCH, FCCH, FCCF, F$_2$CO, H$_2$CO, and HCN taken from the previous CC3 and CCSDT study by Faber et al.\cite{Faber2017_SSCC_CC3_CCSDT}
The geometry optimizations were previously performed at the CCSD(T) level using  the aug-cc-pCVQZ basis set except for the FCCF molecule, where the cc-pCVQZ basis set was used.\cite{Faber2017_SSCC_CC3_CCSDT} 
The spin-spin coupling constants were calculated with the Dalton program.\cite{spas141_dalton20} For every molecule, the SSCCs were calculated with the SOPPA, HRPA(D) and RPA(D)  methods using the aug-ccJ-pVTZ \cite{Benedikt_basis_set-SSCC_aug-ccJ-pvtz}  basis set.

\begin{figure}[h!]
    \centering
    \includegraphics[width=0.5\textwidth]{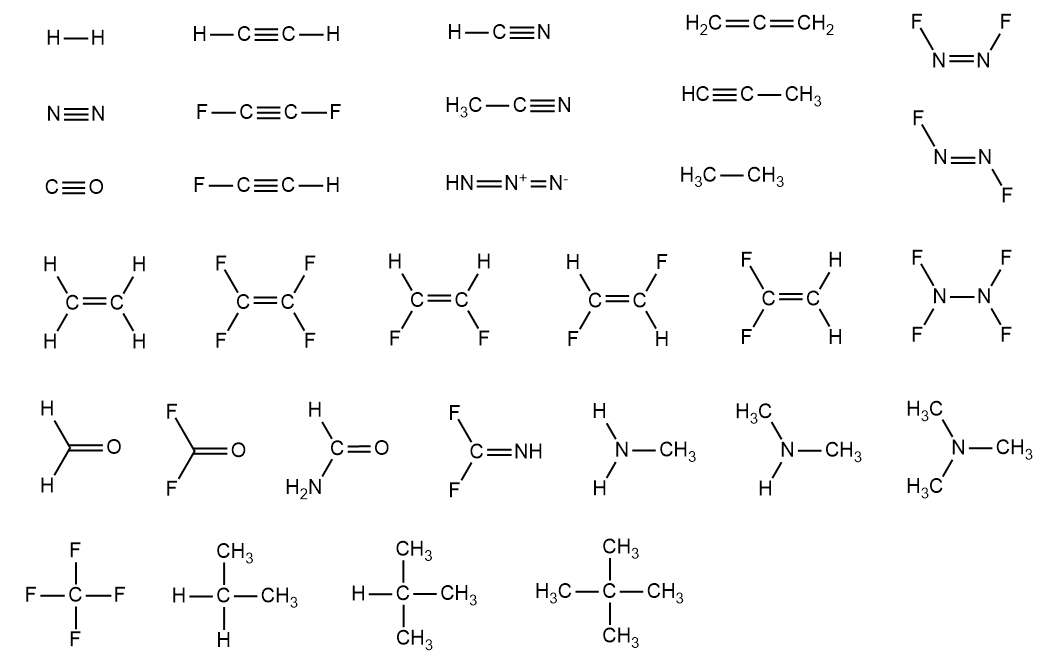}
    \caption{The 31 smaller molecules of set II}
    \label{fig:set2-molecules}
\end{figure}

\textbf{Set II} consists of 31 smaller molecules (see Figure \ref{fig:set2-molecules}).
The geometry optimizations were carried out in this study with the Gaussian program \cite{Gaussian_g16} at the MP2 level\cite{Moller_MP2,Pople_MP2} using the aug-cc-pVTZ basis set.\cite{Dunning1989_basis_set_geo-opt_cc-pvtz,Kendall1992_basis_set_geo-opt_aug-cc-pvtz} 
The calculations of the SSCCs were performed in the same way as for set I. 

\begin{figure}[h!]
    \centering
    \includegraphics[width=0.5\textwidth]{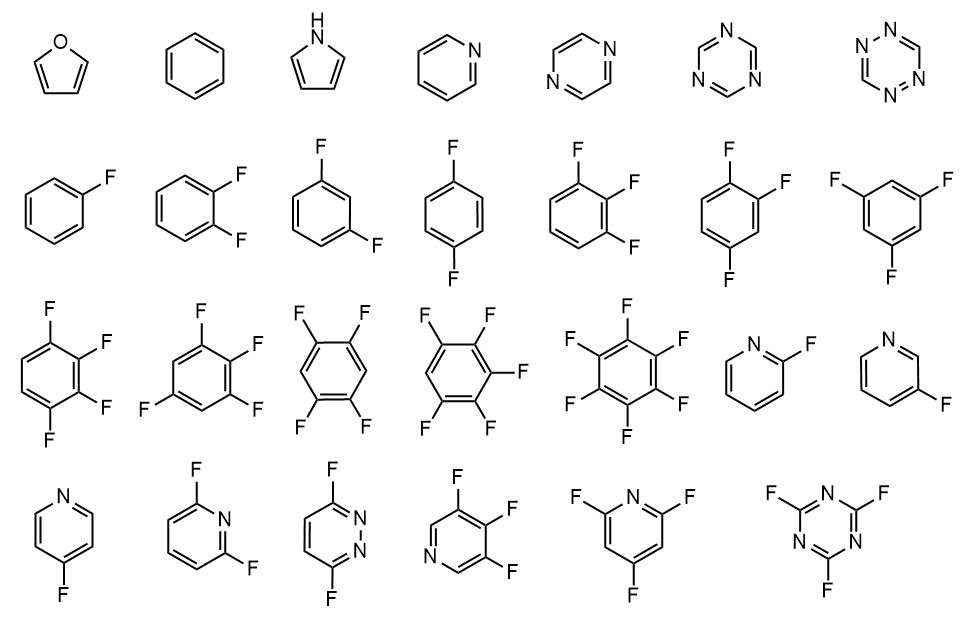}
    \caption{The 27 aromatic and fluoro-aromatic compounds of set III}
    \label{fig:set3-molecules}
\end{figure}

\textbf{Set III} consists of 27 aromatic and fluoro-aromatic compounds (see Figure \ref{fig:set3-molecules}). 
The geometry optimization and SSCCs calculations were carried out in this study in the same way as for set II. 

\section{Results and discussion}
When calculating spin-spin coupling constants, one obtains results for all the components of the coupling tensor. 
However in this study, only the isotropic couplings are reported, since experimental values are mostly used as reference values. 
Some of the reported SSCCs are the mean value of several couplings that experimentally would be found to be indistinguishable due to e.g. rotation. 

A statistical analysis has been performed for every group, which includes the mean absolute deviation (mean abs dev), the mean deviation (mean dev), the standard deviation (std dev), and the maximum absolute deviation (max abs dev).

\subsection{Set I}
For the 7 molecules of set I: CO, HCCH, FCCH, FCCF, F$_2$CO, H$_2$CO, and HCN, Faber et at. \cite{Faber2017_SSCC_CC3_CCSDT} have calculated spin-spin coupling constants at the CC3 and CCSD levels, which will be used as reference values. 
All the SOPPA and HPRA(D) calculated SSCCs together with Faber et al.\cite{Faber2017_SSCC_CC3_CCSDT} CC3 and CCSD values are presented in Table \ref{table:SSCC_geo_CCSDT_noRPAD}. For some of the couplings, it was possible to find experimental values, which are included in the Table as well. 

\begin{table}
\begin{ruledtabular}
\centering
     \caption{All the calculated SSCCs (in Hz) for set I including known experimental values.}
     \label{table:SSCC_geo_CCSDT_noRPAD}
\begin{tabular}{lcrrrrc}
& \textbf{Coupling} & \textbf{CC3$^a$} & \textbf{CCSD$^a$}  &  \textbf{SOPPA} & \textbf{HRPA(D)} &   \textbf{exp$^b$}  \\ \hline
CO	&	$^1J_{CO}$	&	14.62	&	15.07	&	20.26	&	18.40	&	16.4	\cite{Wasylishen1985_exp-value_CO}	\\	\hline
HCCH	&	$^1J_{CH}$	&	240.44	&	243.83	&	261.67	&	260.45	&	248.29		\cite{Kaski1998_exp-value_HCCH-C2H2_H2CCH2-C2H4_H3CCH3-C2H6}	\\	
	&	$^1J_{CC}$	&	180.96	&	187.61	&	190.66	&	188.56	&	169.82		\cite{Kaski1998_exp-value_HCCH-C2H2_H2CCH2-C2H4_H3CCH3-C2H6}	\\	
	&	$^3J_{HH}$	&	9.95	&	10.27	&	11.89	&	11.20	&	9.47		\cite{Kaski1998_exp-value_HCCH-C2H2_H2CCH2-C2H4_H3CCH3-C2H6}	\\	
	&	$^2J_{CH}$	&	53.07	&	51.55	&	52.08	&	50.86	&	49.26		\cite{Kaski1998_exp-value_HCCH-C2H2_H2CCH2-C2H4_H3CCH3-C2H6}	\\	\hline
FCCH	&	$^3J_{FH}$	&	14.45	&	12.72	&	11.61	&	10.28	&	21		\cite{Middleton1959_exp-value_FCCH}	\\	
	&	$^1J_{CF}$	&	-277.68	&	-280.42	&	-305.21	&	-266.28	&		\\			
	&	$^2J_{CF}$	&	25.56	&	26.74	&	22.31	&	23.49	&		\\			
	&	$^1J_{CC}$	&	268.11	&	272.31	&	277.77	&	274.67	&		\\			
	&	$^2J_{CH}$	&	68.53	&	66.84	&	67.85	&	65.63	&		\\			
	&	$^1J_{CH}$	&	270.08	&	272.55	&	292.06	&	289.69	&		\\	\hline		
FCCF	&	$^3J_{FF}$	&	2.56	&	7.26	&	-11.94	&	8.36	&	2.1		\cite{Burger1991_exp-value_FCCF_1}	\\	
	&	$^1J_{CF}$	&	-256.58	&	-260.83	&	-289.18	&	-248.55	&	(-)287.3		\cite{Burger1991_exp-value_FCCF_1}	\\	
	&	$^2J_{CF}$	&	45.54	&	45.14	&	39.85	&	38.24	&	28.7		\cite{Del_Bene2008_exp-value_FCCF_2}	\\	
	&	$^1J_{CC}$	&	401.65	&	407.83	&	417.66	&	410.89	&		\\	\hline		
F$_2$CO	&	$^1J_{CF}$	&	-294.39	&	-292.43	&	-320.82	&	-279.61	&	(-)308		\cite{GOMBLER1981_exp-value_F2CO}	\\	
	&	$^2J_{FF}$	&	-100.20	&	-105.05	&	-120.56	&	-85.35	&		\\			
	&	$^2J_{OF}$	&	39.78	&	38.65	&	41.81	&	32.38	&		\\			
	&	$^1J_{CO}$	&	12.08	&	12.32	&	18.90	&	12.50	&		\\	\hline		
H$_2$CO	&	$^2J_{HH}$	&	37.29	&	36.56	&	40.76	&	38.52	&	40.22		\cite{Shapiro2004_exp-value_H2CO}	\\	
	&	$^2J_{OH}$	&	-3.01	&	-2.82	&	-2.74	&	-3.08	&		\\			
	&	$^1J_{CH}$	&	167.89	&	169.12	&	178.42	&	178.38	&		\\			
	&	$^1J_{CO}$	&	26.98	&	27.18	&	32.34	&	25.49	&		\\	\hline		
HCN	&	$^1J_{CH}$	&	249.95	&	253.82	&	272.40	&	272.74	&	267.3		\cite{FRIESEN1980_exp-value_HCN_2}	\\	
	&	$^1J_{C^{15}N}$	&	-18.19	&	-18.46	&	-15.30	&	-17.96	&	(-)18.5		\cite{FRIESEN1980_exp-value_HCN_2}	\\	
	&	$^2J_{^{15}NH}$	&	-7.47	&	-7.56	&	-7.47	&	-8.31	&	(-)8.7		\cite{Binsch1968_exp-value_HCN_1}	\\
\end{tabular}
\end{ruledtabular}   
    {\raggedright \small $^a$ Calculation performed by Faber et al. \cite{Faber2017_SSCC_CC3_CCSDT}. \\ $^b$ Signs in parentheses are assumed. \par}
\end{table}

\subsubsection{Experimental values as reference}
\begin{figure}[h!]
    \centering
    \includegraphics[width=0.5\textwidth]{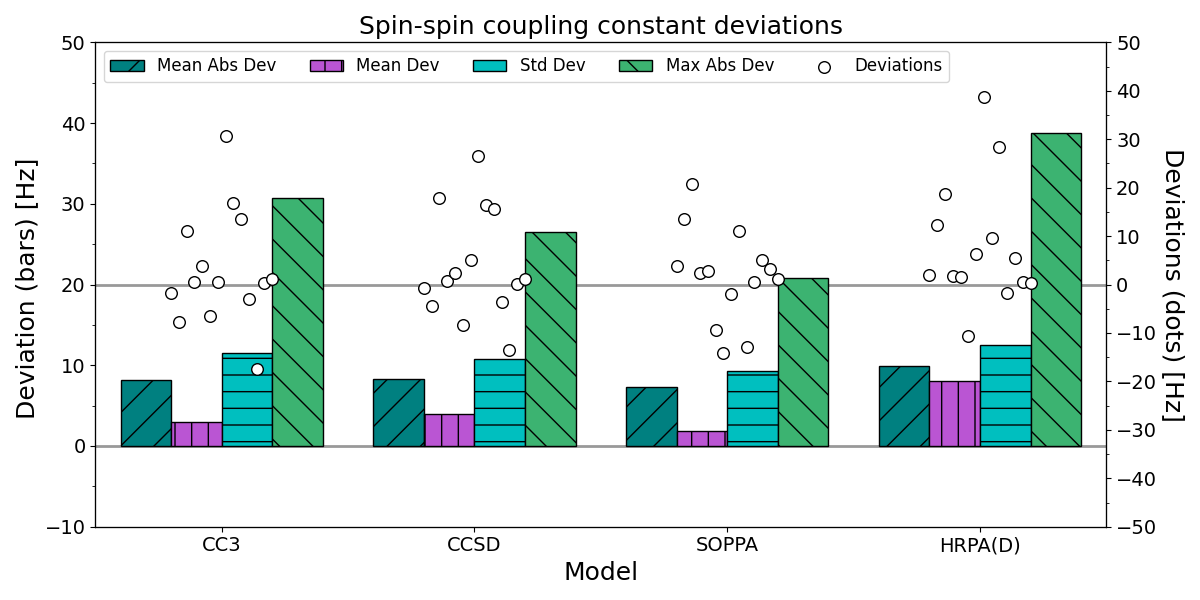}
    \caption{Statistical analysis of the deviations from the experimental values (in Hz) for all methods and all molecules in set I. 
    The dots indicate the deviations of the individual couplings.}
    \label{fig:CCSDT-geo-ref-exp-minusRPAandOF2}
\end{figure}

For the 7 molecules, we are aware of experimental values for 14 SSCCs. 
A statistic analysis has been performed with the experimental values as the reference. 
The results from the statistical analyses are presented in Figure \ref{fig:CCSDT-geo-ref-exp-minusRPAandOF2} and Table \ref{tab:CCSDT-geo_ref-exp}.  

\begin{table}[h!]
\begin{ruledtabular}
\centering
    \caption{Statistical analysis of the deviations from the experimental values (in Hz) for all methods and all molecules in set I. 
    }
    \label{tab:CCSDT-geo_ref-exp}
\begin{tabular}{lrrrr}
 & \textbf{CC3} & \textbf{CCSD} & \textbf{SOPPA} & \textbf{HRPA(D)}   \\ \hline
Count & 14 & 14 & 14 & 14 \\
Mean Abs Dev & 8.21 & 8.30 & 7.34 & 9.86  \\
Mean Dev & 3.02 &	3.94 & 1.89 & 8.09    \\
Std Dev & 11.47 & 10.79 & 9.30 & 12.48    \\
Max Abs Dev & 30.72 & 26.47 & 20.84 & 38.75  \\ 
\end{tabular}
\end{ruledtabular}
\end{table}

From Figure \ref{fig:CCSDT-geo-ref-exp-minusRPAandOF2} and Table \ref{tab:CCSDT-geo_ref-exp}, it can be seen that all the methods mostly overestimate the experimental SSCCs and especially HRPA(D) that only underestimates 2 SSCCs, while SOPPA, CCSD and CC3 underestimate 4, 5, and 5 SSCCs, respectively.
By looking at the consistency (std dev) and accuracy (mean abs dev) of the calculated SSCCs, SOPPA seems to perform better than all the other methods followed by CC3.
HRPA(D) performs quite well, it manages to perform closely to CC3 and CCSD.
The coupling constant deviating mostly from experimental values are for CC3, CCSD, and HRPA(D) the $^1J_{CF}(\text{FCCF})$ coupling, and for SOPPA it is the $^1J_{CC}(\text{HCCH})$ coupling. 

Even though CC3 is the most precise and expensive method of all the methods used in the present study, employing it in equilibrium geometry gas phase calculations results not necessarily in the best agreement with experimental SSCCs as already noted by Faber et at. \cite{Faber2017_SSCC_CC3_CCSDT} 
This is both notable from the statistical analysis and when looking at every SSCC separately. 

For 5 out of the 14 SSCCs, CC3 deviates more than 10 Hz from the experimental values, which are the coupling constants $^1J_{CC}(\text{HCCH})$, $^1J_{CF}(\text{FCCF})$, $^2J_{CF}(\text{FCCF})$, $^1J_{CF}(\text{F}_2\text{CO})$, and $^1J_{CH}(\text{HCN})$. For these (except $^1J_{CC}(\text{HCCH})$) either both SOPPA and HRPA(D) or only SOPPA perform better than CC3. 

For 4 out of 14 SSCCs, CC3 performs better than the other methods. The 4 coupling constants are $^1J_{CC}(\text{HCCH})$, $^3J_{HH}(\text{HCCH})$, $^3J_{FH}(\text{FCCH})$ and $^3J_{FF}(\text{FCCF})$. 
For $^1J_{CC}(\text{HCCH})$, all the methods overestimate the value with a deviation of 11.14 Hz for CC3 and deviations between 17.79 and 20.84 Hz for CCSD, HRPA(D), and SOPPA. 
For 2 out of 14 SSCCs ($^1J_{CH}(\text{HCCH})$ and $^1J_{CN}(\text{HCN})$.), CCSD performs best followed by CC3. 

For the 4 SSCCs $^3J_{FF}(\text{FCCF})$, $^1J_{CF}(\text{FCCF})$, $^1J_{CF}(\text{F}_2\text{CO})$, and $^2J_{HH}(\text{H}_2\text{CO})$, the deviation from experimental values for SOPPA has the opposite signs than the other methods.
For the $^1J_{CF}(\text{FCCF})$ coupling, SOPPA performs much better than all the other methods in the present study, where SOPPA underestimates the value with 1.88 Hz and the other methods all overestimate it with 26 to 39 Hz.  

For the 3 SSCCs $^1J_{CO}(\text{CO})$, $^1J_{CH}(\text{HCCH})$, and $^1J_{CH}(\text{HCN})$, the deviation from experimental values for SOPPA and HRPA(D) have the opposite signs than the ones for CC3 and CCSD, where SOPPA and HRPA(D) overestimate and CC3 and CCSD underestimates the values. 
For $^1J_{CO}(\text{CO})$, the absolute deviations from experimental values are within 3.86 Hz for all four methods.

For the $^2J_{NH}(\text{HCN})$ coupling, all the methods overestimate the value slightly with the deviations from experimental values lying within 0.84 Hz of each other with HRPA(D) performing best with a deviation at 0.39 Hz. 

When comparing SOPPA and HRPA(D) to each other, 11 out of the 14 SSCCs are within 3 Hz of each other. The 3 coupling constants that differ more are $^3J_{FF}(\text{FCCF})$ (differ with 20.30 Hz), $^1J_{CF}(\text{FCCF})$ (differ with 40.63 Hz), and $^1J_{CF}(\text{F}_2\text{CO})$ (differ with 41.21 Hz). For these 3 coupling constants, SOPPA underestimates and HRPA(D) overestimates the value.

The deviation from experimental values can be due to the calculation not including solvent effects and vibrational averaging. 
Therefore, it would make sense to try to use the CC3 values as references, since they should exhibit the same calculation error as with the other methods. This is investigated in the next section.

\subsubsection{CC3 results as reference}
\begin{figure}[h!]
    \centering
    \includegraphics[width=0.5\textwidth]{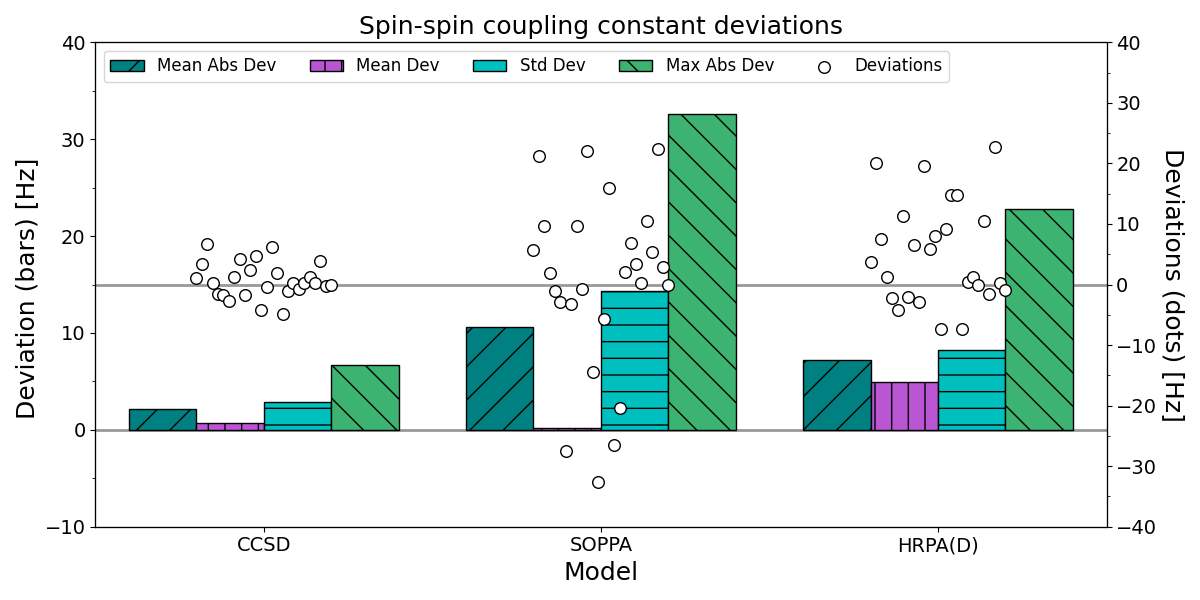}
    \caption{Statistical analysis of the deviations from the CC3 values (in Hz) for all methods and all molecules in set I. 
    The dots indicated the deviations of the individual couplings.}
    \label{fig:CCSDT-geo-ref-CC3}
\end{figure}

For the 7 molecules, a total of 26 SSCCs have been calculated. 
A statistic analysis has been performed with the CC3 SSCCs as the reference. 
The results for the analyses are presented in Figure \ref{fig:CCSDT-geo-ref-CC3} and Table \ref{tab:CCSDT-geo_ref-CC3}.

\begin{table}[h!]
\centering
    \caption{Statistical analysis of the deviations from the CC3 values (in Hz) for all methods and all molecules in set I.}
    \label{tab:CCSDT-geo_ref-CC3}
\begin{tabular}{lrrr}
\hline
& \textbf{CCSD} & \textbf{SOPPA} & \textbf{HRPA(D)}   \\ \hline
 Count & 26 & 26 & 26 \\
 Mean Abs Dev & 2.20 & 10.57 & 7.17   \\
 Mean Dev & 0.71 & 0.20 & 4.99     \\
 Std Dev & 2.84 & 14.31 & 8.29     \\
 Max Abs Dev & 6.65 & 32.60 & 22.79   \\ \hline
\end{tabular}
\end{table}

When using the CC3 values as references, CCSD does not surprisingly perform best with the lowest value of accuracy and consistency. Of SOPPA and HRPA(D), HRPA(D) performs considerably better than SOPPA. 
The SSCCs which result in the largest deviation from CC3 are different for all the methods. For CCSD, the coupling constant $^1J_{CC}(\text{HCCH})$ is overestimated with 6.65 Hz. For SOPPA, the coupling constant $^1J_{CF}(\text{FCCF})$ is underestimated with 32.60 Hz. For HRPA(D), the coupling constant $^1J_{CH}(\text{HCN})$ is overestimated with 22.79 Hz.

For 8 out of 26 SSCCs, CCSD clearly performs better than both SOPPA and HRPA(D) with a difference between them being over 7 Hz and sometimes over 20 Hz. 
The coupling constants are $^1J_{CH}(\text{HCCH})$, $^1J_{CF}(\text{FCCH})$, 
$^1J_{CH}(\text{FCCH})$, $^1J_{CF}(\text{FCCF})$, $^1J_{CF}(\text{F}_2\text{CO})$, $^2J_{FF}(\text{F}_2\text{CO})$, $^1J_{CH}(\text{H}_2\text{CO})$, and $^1J_{CH}(\text{HCN})$. 
For the coupling constant $^3J_{FF}(\text{FCCF})$, CCSD performs clearly better than SOPPA which underestimates the value with 14.50 Hz, whereas HRPA(D) overestimates the values with 1.10 Hz more than the CCSD SSCC which overestimates it with 4.70 Hz. 

For most of the SSCCs, CCSD is closer to the CC3 values than SOPPA and HRPA(D) are. 
However, there are a few SSCCs where either SOPPA or HRPA(D) perform slightly better. 
For the 2 out of 26 SSCCs, $^2J_{CH}(\text{HCCH})$ and $^2J_{CH}(\text{FCCH})$, the SOPPA value are closer to CC3, and for $^2J_{NH}(\text{HCN})$ they are identical. 
For the 2 out of 26 SSCCs, $^2J_{OH}(\text{H}_2\text{CO})$ and $^1J_{CN}(\text{HCN})$, HRPA(D) performs closer to the CC3 value than the other methods. 
Even though SOPPA or HRPA(D) are performing better for these coupling 5 couplings, the CCSD SSCCs are within 1 Hz of the SOPPA/HRPA(D) SSCCs.

When comparing the SSCCs of SOPPA and HRPA(D), 21 out of 26 SSCCs are within 10 Hz of each other. The 5 SSCCs that differ more are $^1J_{CF}$ for FCCH, FCCF, and F$_2$CO that differ by 38.93 Hz, 40.63 Hz, and 41.21 Hz respectively, $^3J_{FF}(\text{FCCF})$ that differ by 20.30 Hz, and $^2J_{FF}(\text{F}_2\text{CO})$ that differ by 35.21 Hz.
For most of the SSCCs (17 out of 26), SOPPA and HRPA(D) perform within 5 Hz of each other. 

For four coupling constants, the deviation from CC3 for SOPPA and HRPA(D) have opposite signs. 
For two of these coupling constants ($^1J_{CF}(\text{F}_2\text{CO})$ and $^2J_{FF}(\text{F}_2\text{CO})$), SOPPA largely underestimates the values with 26.43 Hz and 20.36 Hz, respectively, whereas HRPA(D) overestimates the values with 14.78 Hz and 14.85 Hz, respectively. 
For the last two coupling constants ($^2J_{OF}(\text{F}_2\text{CO})$ and $^1J_{CO}(\text{H}_2\text{CO})$), HRPA(D) underestimates the values with 7.40 Hz and 1.49 Hz, respectively, and SOPPA overestimates them with 2.03 Hz and 5.36 Hz. However, the difference between SOPPA and HRPA(D) for these two coupling constants are 9.43 Hz and 6.85 Hz, and therefore not considered abnormal. 

CCSD performs better than SOPPA and HRPA(D) when CC3 SSCCs are used as the reference values, however, the method is also more expensive. HRPA(D) performs significantly better than SOPPA and would be an adequate option since it is cheaper. 

\subsubsection{Effects of the geometry optimization for set I}
Several authors have previously investigated how different geometries used for SSCC calculations affect the values.
Helgaker et al.\cite{Helgaker_mix-in-geo-opt-and-sscc-calc} investigated (among other things) how the geometry affected the SSCCs calculation for o-benzyne, where the calculations were performed from an experimental geometry and an optimized geometry. The SSCCs with an optimized geometry were closer to the experimental values. 
Most recently, Giovanetti et al.\cite{Marinella-Giovanetti_mix-in-geo-opt-and-sscc-calc} investigated how a mix in level of theory for the geometry optimization and in SSCCs calculation affected the $^1J_{FC}$ SSCCs. The results show that the SSCCs, where the same functional was used in both the geometry optimization and the SSCC calculation, performed most of the time more accurately.
The conclusion of these studies is that using the same level of theory in both the geometry optimization and the SSCCs calculation decreases the triplet instabilities, and therefore results in more precise SSCCs.  

In the present study, we have therefore investigated whether a CCSD(T) or a MP2 geometry optimization performs best for SOPPA and HRPA(D) SSCC calculation compared with experimental values. 
This is possible since the 7 molecules of set I (CO, HCCH, FCCH, FCCF, F$_2$CO, H$_2$CO, and HCN) are part of set II as well, and therefore the molecules have been optimized with both CCSD(T) and MP2. 
These SSCCs calculated for set I at both MP2 and CCSD(T) geometries are presented together in Table \ref{table:SSCC_mixed_geo}.

\begin{table}[h!]
    \caption{Comparision of the calculated SSCCs of set I at both CCSD(T) and MP2 optimized geometries.}
    \label{table:SSCC_mixed_geo}
\centering
\begin{tabular}{lclrrc}
\hline
\textbf{Molecule} &  \textbf{Coupling} & \textbf{Geo-opt} &  \textbf{SOPPA} &  \textbf{HRPA(D)}  &  \textbf{exp$^a$}  \\ \hline
CO & $^1J_{CO}$ & CCSD(T) & 20.26 & 18.40 &  16.4  \cite{Wasylishen1985_exp-value_CO} \\
 &  & MP2 & 23.05 & 21.08 &  \\ \hline
HCCH & $^1J_{CH}$ & CCSD(T) & 261.67 & 260.45 &  248.29  \cite{Kaski1998_exp-value_HCCH-C2H2_H2CCH2-C2H4_H3CCH3-C2H6} \\ 
 &  & MP2 & 263.96 & 262.67 &   \\ \cline{2-6} 
 & $^1J_{CC}$ & CCSD(T) & 190.66 & 188.56 &  169.82  \cite{Kaski1998_exp-value_HCCH-C2H2_H2CCH2-C2H4_H3CCH3-C2H6} \\ 
 &  & MP2 & 190.00 & 187.80 &   \\ \cline{2-6} 
 & $^3J_{HH}$ & CCSD(T) & 11.89 & 11.20 &  9.47  \cite{Kaski1998_exp-value_HCCH-C2H2_H2CCH2-C2H4_H3CCH3-C2H6} \\ 
 &  & MP2 & 12.24 & 11.52 &   \\ \cline{2-6} 
 & $^2J_{CH}$ & CCSD(T) & 52.08 & 50.86 &  49.26 \cite{Kaski1998_exp-value_HCCH-C2H2_H2CCH2-C2H4_H3CCH3-C2H6} \\
 &  & MP2 & 51.80 & 50.59 &   \\ \hline
FCCH & $^2J_{FH}$ & CCSD(T) & 11.61 & 10.28 & 21  \cite{Middleton1959_exp-value_FCCH} \\
 &  & MP2 & 11.61 & 10.31 &   \\ \hline
FCCF & $^3J_{FF}$ & CCSD(T) & -11.94 & 8.36  & 2.1   \cite{Burger1991_exp-value_FCCF_1} \\
 &  & MP2 & -19.62 & 4.04  &  \\ \cline{2-6} 
 & $^1J_{CF}$ & CCSD(T) & -289.18 & -248.55  & (-)287.3  \cite{Burger1991_exp-value_FCCF_1} \\
 &  & MP2 & -299.44 & -257.67 &   \\ \cline{2-6} 
 & $^2J_{CF}$ & CCSD(T) & 39.85 & 38.24  & 28.7 \cite{Del_Bene2008_exp-value_FCCF_2} \\
 &  & MP2 & 39.49 & 37.64 &   \\ \hline
F$_2$CO & $^1J_{CF}$ & CCSD(T) & -320.82 & -279.61 &  (-)308  \cite{GOMBLER1981_exp-value_F2CO} \\
 &  & MP2 & -330.28 & -287.96 &   \\ \hline
H$_2$CO & $^2J_{HH}$ & CCSD(T) & 40.76 & 38.52 &  40.22  \cite{Shapiro2004_exp-value_H2CO} \\
 &  & MP2 & 39.93 & 37.74 &   \\ \hline
HCN & $^1J_{CH}$ & CCSD(T) & 272.40 & 272.74 &  267.3  \cite{FRIESEN1980_exp-value_HCN_2} \\
 &  & MP2 & 273.39 & 273.70 &  \\ \cline{2-6} 
 & $^1J_{CN}$ & CCSD(T) & -15.30 & -17.96 &  (-)18.5 \cite{FRIESEN1980_exp-value_HCN_2} \\ 
 &  & MP2 & -13.83 & -16.50 &   \\ \cline{2-6} 
 & $^2J_{NH}$ & CCSD(T) & -7.47 & -8.31 &  (-)8.7  \cite{Binsch1968_exp-value_HCN_1} \\
 &  & MP2 & -7.03 & -7.89 &   \\ \hline
\end{tabular}
    {\raggedright \\ $^a$ Signs in parentheses are assumed. \par}
\end{table}

In general, the differences between the SSCCs of the two geometry optimization are not large and mostly they are within 3 Hz from each other, except for the three coupling constants $^3J_{FF}(\text{FCCF})$ (differ 7.68 Hz for SOPPA and 4.32 Hz for HRPA(D)), $^1J_{CF}(\text{FCCF})$ (differ 10.26 Hz for SOPPA and 9.12 Hz for HRPA(D)), and $^1J_{CF}(\text{F}_2\text{CO})$ (differ 9.46 Hz for SOPPA and 8.35 Hz for HRPA(D). 

A statistical analysis has been performed with the experimental values used as references. The results of the analysis are presented for the SSCCs calculation with both geometry optimizations in Table \ref{tab:mixed-geo_ref-exp}. 

\begin{table}[h!]
    \caption{Statistical analysis of the deviation from the experimental values (in Hz) for the 7 molecules (14 SSCCs) in set I calculated at CCSD(T) and MP2 optimized geometries.} 
    \label{tab:mixed-geo_ref-exp}
\centering
\begin{tabular}{lrrrr}
\hline
 & \textbf{Mean Abs} & \textbf{Mean Dev} & \textbf{Std Dev} & \textbf{Max Abs} \\ 
  & \textbf{Dev} & & & \textbf{Dev} \\ \hline
\textbf{SOPPA//CCSD(T)} & 7.33 & 1.89 & 9.29 & 20.84 \\
\textbf{SOPPA//MP2} & 9.77 & 0.37 & 12.22 & 22.28 \\ \hline
\textbf{HRPA(D)//CCSD(T)} & 9.85 & 8.08 & 12.48 & 38.75 \\
\textbf{HRPA(D)//MP2} & 8.81 & 6.93 & 10.05 & 29.63 \\ \hline
\end{tabular}
\end{table}

For the 7 molecules, the two methods SOPPA and HRPA(D) prefer different methods for the geometry optimization. 
For the SOPPA SSCCs, the CCSD(T) geometry optimization results in slightly better values than the MP2 geometry optimization. 
For HRPA(D), a MP2 geometry results in slightly better SSCCs than the CCSD(T) geometry. 

It was expected that the CCSD(T) geometry optimization would perform best since it is a more precise and expensive method. 
In general, a CCSD(T) geometry would be closer to the experimental geometry. 
However, it does not necessarily mean that it gives the best geometry for the calculation of SSCCs, which is the case for the SSCCs calculated with HRPA(D) in the present study. 
Likewise, an optimized geometry performed best in the DFT study of Helgaker et al.\cite{Helgaker_mix-in-geo-opt-and-sscc-calc} compared to an experimental geometry.

\subsection{Set II: Smaller molecules}
For the 31 molecules of set II, 72 different spin-spin coupling constants have been calculated with the methods SOPPA and HPRA(D). 
All the SSCCs, including the experimental values, can be found in Table \ref{table:all-SSCC_set2_noRPA}.

\begin{longtable}{llrrc}
\caption{The calculated SSCCs in set II including the experimental values given in Hz.}
    \label{table:all-SSCC_set2_noRPA} \\
\hline
\textbf{Molecule} & \textbf{SSCC} & \textbf{SOPPA} & \textbf{HRPA(D)} &  \textbf{exp$^a$} \\
\hline
\endfirsthead
\multicolumn{5}{c}%
{\tablename\ \thetable\ -- \textit{Continued from previous page}} \\
\hline
\textbf{Molecule} & \textbf{SSCC} & \textbf{SOPPA} & \textbf{HRPA(D)} &  \textbf{exp$^a$}  \\
\hline
\endhead
\hline \multicolumn{5}{r}{\textit{Continued on next page}} \\
\endfoot
\hline
    \caption*{$^a$ Signs in parentheses are assumed. \\ $^b$ from $^1J_{HD}$ 42.94 Hz.}
\endlastfoot
CO	&	$^1J_{CO}$	&	23.05	&	21.08	&	16.4	\cite{Wasylishen1985_exp-value_CO}	\\ \hline
H$_2$	&	$^1J_{HH}$	&	286.67	&	286.54		&	279.73$^{b,}$	\cite{Beckett1981_exp-value_H2,oddershede1988_exp-value_H2}	\\ \hline
N$_2$	&	$^1J_{^{14}N^{15}N}$	&	3.81	&	3.33	&	1.8	\cite{Friedrich1985_exp-value_N2}	\\ \hline
N$_3$H	&	$^1J_{^{15}N2^{15}N3}$	&	-17.39	&	-14.52	&	(-)13.95		\cite{book_Berger1997_non-metallic-elements-NMR_exp-values}	\\
	&	$^1J_{^{15}N1^{15}N2}$ &	-10.60	&	-7.77	&	(-)7.2		\cite{book_Berger1997_non-metallic-elements-NMR_exp-values}	\\
	&	$^1J_{^{15}NH}$	&	-69.20	&	-72.45	&	-70.2		\cite{book_Berger1997_non-metallic-elements-NMR_exp-values}	\\ \hline
HCCH	&	$^1J_{CH}$	&	263.96	&	262.67	&	248.29		\cite{Kaski1998_exp-value_HCCH-C2H2_H2CCH2-C2H4_H3CCH3-C2H6}	\\
	&	$^1J_{CC}$	&	190.00	&	187.80	&	169.82		\cite{Kaski1998_exp-value_HCCH-C2H2_H2CCH2-C2H4_H3CCH3-C2H6}	\\
 	&	$^3J_{HH}$	&	12.24	&	11.52	&	9.47		\cite{Kaski1998_exp-value_HCCH-C2H2_H2CCH2-C2H4_H3CCH3-C2H6} \\
	&	$^2J_{CH}$	&	51.80	&	50.59	&	49.26		\cite{Kaski1998_exp-value_HCCH-C2H2_H2CCH2-C2H4_H3CCH3-C2H6}	\\ \hline
FCCH	&	$^3J_{FH}$	&	11.61	&	10.31	&	21		\cite{Middleton1959_exp-value_FCCH}	\\ \hline
FCCF	&	$^3J_{FF}$	&	-19.62	&	4.04	&	2.1		\cite{Burger1991_exp-value_FCCF_1}	\\
	&	$^1J_{CF}$	&	-299.44	&	-257.67	&	(-)287.3		\cite{Burger1991_exp-value_FCCF_1}	\\
	&	$^2J_{CF}$	&	39.49	&	37.64	&	28.7		\cite{Del_Bene2008_exp-value_FCCF_2}	\\ \hline
F$_2$CO	&	$^1J_{CF}$	&	-330.28	&	-287.96	&	(-)308		\cite{GOMBLER1981_exp-value_F2CO}	\\ \hline
H$_2$CO	&	$^2J_{HH}$	&	39.93	&	37.74	&	40.22		\cite{Shapiro2004_exp-value_H2CO}	\\ \hline
HCN	&	$^1J_{CH}$	&	273.39	&	273.70	&	267.3		\cite{FRIESEN1980_exp-value_HCN_2}	\\
	&	$^1J_{C^{15}N}$	&	-13.83	&	-16.50	&	(-)18.5		\cite{FRIESEN1980_exp-value_HCN_2}	\\
	&	$^2J_{^{15}NH}$	&	-7.03	&	-7.89	&	(-)8.7		\cite{Binsch1968_exp-value_HCN_1}	\\ \hline
H$_3$CC$\equiv$N	&	$^1J_{CH}$	&	136.35	&	136.01	&	136.27		\cite{WILCZEK2002_exp-value_CH3CN}	\\
	&	$^2J_{CH}$	&	-11.88	&	-11.35	&	-9.74		\cite{WILCZEK2002_exp-value_CH3CN}	\\
	&	$^1J_{CC}$	&	63.52	&	65.55	&	56.99		\cite{WILCZEK2002_exp-value_CH3CN}	\\
	&	$^1J_{C^{15}N}$	&	-12.73	&	-15.96	&	-17.55		\cite{WILCZEK2002_exp-value_CH3CN}	\\
	&	$^2J_{C^{15}N}$	&	3.61	&	2.51	&	2.81		\cite{WILCZEK2002_exp-value_CH3CN}	\\
	&	$^3J_{^{15}NH}$	&	-1.74	&	-1.51 &	-1.70		\cite{WILCZEK2002_exp-value_CH3CN}	\\ \hline
H$_2$C=CH$_2$	&	$^1J_{CC}$	&	71.02	&	71.14	&	67.54		\cite{Kaski1998_exp-value_HCCH-C2H2_H2CCH2-C2H4_H3CCH3-C2H6}	\\
	&	$^1J_{CH}$	&	160.93	&	160.46	&	156.30		\cite{Kaski1998_exp-value_HCCH-C2H2_H2CCH2-C2H4_H3CCH3-C2H6}	\\ 
 	&	$^3J_{HH}$ \small cis	&	12.74	&	12.13	&	11.62		\cite{Kaski1998_exp-value_HCCH-C2H2_H2CCH2-C2H4_H3CCH3-C2H6}	\\	
	&	$^3J_{HH}$ \small trans	&	19.09	&	18.20	&	19.02		\cite{Kaski1998_exp-value_HCCH-C2H2_H2CCH2-C2H4_H3CCH3-C2H6}	\\	\hline
F$_2$C=CF$_2$	&	$^2J_{FF}$	&	107.17	&	105.97	&	124		\cite{book_Berger1997_non-metallic-elements-NMR_exp-values}	\\
	&	$^3J_{FF}$ \small cis	&	81.72	&	51.73	&	73.3		\cite{book_Berger1997_non-metallic-elements-NMR_exp-values}	\\
	&	$^3J_{FF}$ \small trans	&	-117.33	&	-89.19	&	-114		\cite{book_Berger1997_non-metallic-elements-NMR_exp-values} \\ \hline
F$_2$C=CH$_2$	&	$^1J_{CF}$	&	-301.28	&	-266.04	&	(-)287		\cite{book_kalinowski1988_carbon13-NMR-spec_exp-values}	\\
	&	$^2J_{FF}$	&	14.28	&	40.53	&	32.7		\cite{book_Berger1997_non-metallic-elements-NMR_exp-values}	\\ \hline
trans-FHC=CHF 	&	$^3J_{FF}$	&	-139.49	&	-103.15	&	-132.7		\cite{book_Berger1997_non-metallic-elements-NMR_exp-values}	\\ \hline
cis-FHC=CHF 	&	$^2J_{CF}$	&	9.57	&	6.16	&	5.9		\cite{book_kalinowski1988_carbon13-NMR-spec_exp-values}	\\
	&	$^3J_{FF}$	&	-14.42	&	-15.25	&	-18.7		\cite{book_Berger1997_non-metallic-elements-NMR_exp-values}	\\ \hline
trans-FNNF	&	$^1J_{^{15}NF}$	&	193.38	&	170.18	&	172.8		\cite{book_Berger1997_non-metallic-elements-NMR_exp-values}	\\
	&	$^2J_{^{15}NF}$	&	-58.93	&	-49.63	&	-62.8		\cite{book_Berger1997_non-metallic-elements-NMR_exp-values}	\\ \hline
cis-FNNF	&	$^1J_{^{15}NF}$	&	233.08	&	199.20	&	211.0		\cite{book_Berger1997_non-metallic-elements-NMR_exp-values}	\\
	&	$^2J_{^{15}NF}$	&	-13.37	&	-5.07	&	-25.4		\cite{book_Berger1997_non-metallic-elements-NMR_exp-values}	\\ \hline
F$_2$NNF$_2$	&	$^1J_{^{15}NF}$	&	179.62	&	165.03	&	164		\cite{book_Berger1997_non-metallic-elements-NMR_exp-values}	\\ \hline
CF$_4$	&	$^1J_{CF}$	&	-272.51	&	-241.62	&	(-)259.2		\cite{book_kalinowski1988_carbon13-NMR-spec_exp-values}	\\ \hline
F$_2$CNH	&	$^2J_{FF}$	&	-69.29	&	-32.47	&	-54.6	\cite{book_Berger1997_non-metallic-elements-NMR_exp-values}	\\ \hline
H$_3$CC$\equiv$CH	&	$^1J_{CC}$	&	73.73	&	74.37	&	67.4		\cite{book_kalinowski1988_carbon13-NMR-spec_exp-values,book_stothers1972_carbon13-NMR-spec_exp-values}	\\	
	&	$^1J_{CC}$ \small ($\equiv$)	&	190.81	&	188.70	&	175.0	\cite{book_kalinowski1988_carbon13-NMR-spec_exp-values}	\\	
	&	$^2J_{CC}$	&	12.50	&	12.48	&	11.8	\cite{book_stothers1972_carbon13-NMR-spec_exp-values}	\\	
	&	$^1J_{CH}$ \small (Me)	&	133.05	&	132.70	&	131.6	\cite{book_kalinowski1988_carbon13-NMR-spec_exp-values}	\\	
	&	$^1J_{CH}$ \small (CH)	&	263.14	&	261.70	&	248.1	\cite{book_kalinowski1988_carbon13-NMR-spec_exp-values}	\\	\hline
H$_2$C=C=CH$_2$	&	$^1J_{CH}$	&	174.45	&	172.55	&	167.8	\cite{book_kalinowski1988_carbon13-NMR-spec_exp-values}	\\	
	&	$^1J_{CC}$	&	104.95	&	104.72	&	98.7	\cite{book_kalinowski1988_carbon13-NMR-spec_exp-values}	\\	\hline
H$_3$CCH$_3$	&	$^1J_{CH}$	&	126.37	&	126.41	&	125.19	\cite{Kaski1998_exp-value_HCCH-C2H2_H2CCH2-C2H4_H3CCH3-C2H6}	\\	
	&	$^1J_{CC}$	&	36.25	&	36.89	&	34.56	\cite{Kaski1998_exp-value_HCCH-C2H2_H2CCH2-C2H4_H3CCH3-C2H6}	\\	\hline
H$_2$C(CH$_3$)$_2$	&	$^1J_{CH}$ \small (Me)	&	125.76	&	125.79	&	124.4	\cite{book_kalinowski1988_carbon13-NMR-spec_exp-values}	\\	
	&	$^1J_{CH}$	&	126.92	&	126.95	&	125.4	\cite{book_kalinowski1988_carbon13-NMR-spec_exp-values}	\\	
	&	$^1J_{CC}$	&	36.61	&	37.30	&	34.6	\cite{book_kalinowski1988_carbon13-NMR-spec_exp-values}	\\	\hline
HC(CH$_3$)$_3$	&	$^1J_{CH}$ \small (Me)	&	125.54	&	125.56	&	124.0		\cite{book_kalinowski1988_carbon13-NMR-spec_exp-values}	\\	\hline
C(CH$_3$)$_4$	&	$^1J_{CH}$ \small (Me)	&	125.60	&	125.62	&	123.3		\cite{book_kalinowski1988_carbon13-NMR-spec_exp-values}	\\	
	&	$^1J_{CC}$	&	37.37	&	38.17	&	36.9		\cite{book_stothers1972_carbon13-NMR-spec_exp-values}	\\	\hline
NH$_2$(CH$_3$)	&	$^1J_{^{15}NH}$	&	-66.66	&	-67.32	&	-64.5		\cite{book_Berger1997_non-metallic-elements-NMR_exp-values}	\\	
	&	$^1J_{CH}$	&	133.61	&	133.65	&	133		\cite{book_kalinowski1988_carbon13-NMR-spec_exp-values}	\\	\hline
NH(CH$_3$)$_2$	&	$^1J_{^{15}NH}$	&	-69.34	&	-70.07	&	-67.0		\cite{book_Berger1997_non-metallic-elements-NMR_exp-values}	\\	\hline
N(CH3)$_3$	&	$^1J_{CH}$	&	135.91	&	135.80	&	133		\cite{book_kalinowski1988_carbon13-NMR-spec_exp-values}	\\	\hline
H$_2$NCHO	&	$^1J_{C^{14}N}$	&	11.25	&	12.38	&	10.57		\cite{Vaara1997_formamide_H2NCHO_14-N}	\\	
	&	$^1J_{CH}$	&	193.66	&	192.96	&	193.11		\cite{Vaara1997_formamide_H2NCHO_14-N}	\\	
	&	$^3J_{HH}$	&	0.79	&	0.88	&	2.28		\cite{Sorensen1981_formamides_H2NCHO_15-N}	\\	
	&	$^3J_{HH}$	&	12.73	&	12.30	&	12.92		\cite{Sorensen1981_formamides_H2NCHO_15-N}	\\	
	&	$^2J_{HH}$	&	4.28	&	4.17	&	2.60		 \cite{Sorensen1981_formamides_H2NCHO_15-N}	\\	
	&	$^1J_{^{15}NH}$	&	-94.68	&	-94.69	&	-87.80		\cite{Sorensen1981_formamides_H2NCHO_15-N}	\\	
	&	$^1J_{^{14}NH}$	&	67.50	&	67.50	&	63.23		\cite{Vaara1997_formamide_H2NCHO_14-N}	\\	
	&	$^2J_{^{15}NH}$	&	-18.77	&	-17.09	&	-14.25		\cite{Sorensen1981_formamides_H2NCHO_15-N}	\\	
	&	$^2J_{^{14}NH}$	&	13.38	&	12.18	&	9.62		\cite{Vaara1997_formamide_H2NCHO_14-N}	\\	\hline
\end{longtable}

\begin{figure}[h!]
    \centering
    \includegraphics[width=0.5\textwidth]{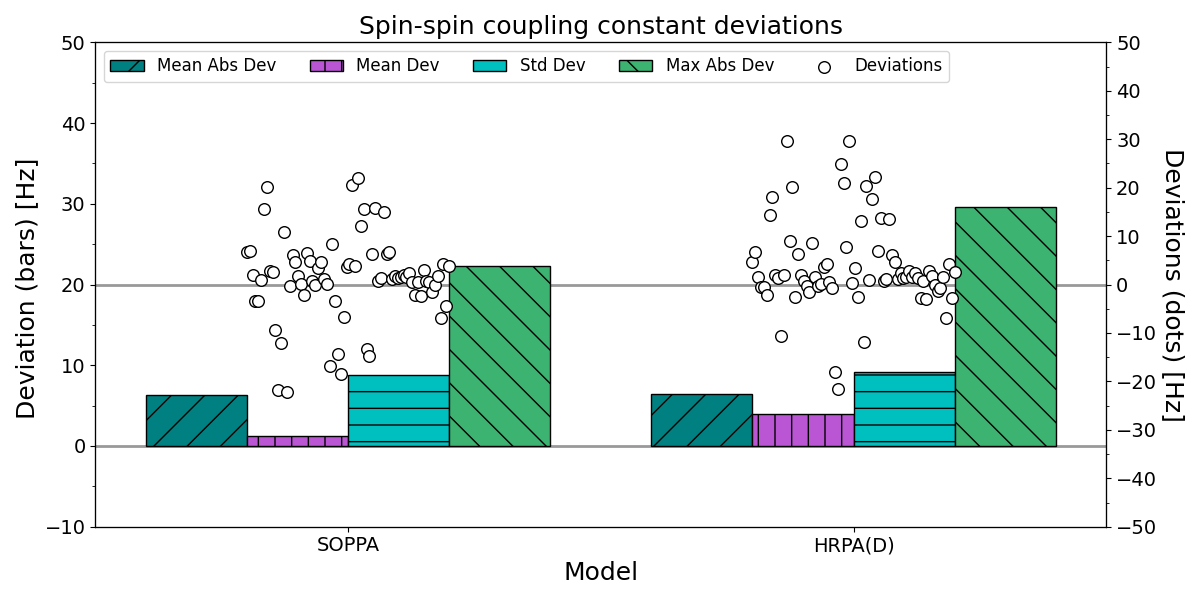}
    \caption{Statistical analysis of the deviations from the experimental values (in Hz) for SOPPA and HRPA(D) and all molecules in set II. 
    The dots indicated the deviations of the individual couplings.}
    \label{fig:MP2-geo-minusRPAandOF2}
\end{figure}

A statistical analysis has been performed with the experimental values as reference, the results of which are presented in Figure \ref{fig:MP2-geo-minusRPAandOF2} and Table \ref{table:Statistics_All_SSCC_MP2-geo_small_molecules}.

\begin{table}[h!]
    \caption{Statistical analysis of the deviations from the experimental values (in Hz) for all methods and all molecules in set II.}
    \label{table:Statistics_All_SSCC_MP2-geo_small_molecules}
\centering
\begin{tabular}{lrr}
\hline
& \textbf{SOPPA} & \textbf{HRPA(D)}  \\ \hline
 Count & 72 & 72 \\  
 Mean Abs Dev & 6.28 & 6.45  \\
 Mean Dev & 1.27 & 3.91  \\
 Std Dev & 8.82 & 9.20  \\
 Max Abs Dev & 22.28 & 29.63  \\ 
\hline
\end{tabular}
\end{table}

From the analysis, it becomes clear that SOPPA performs in general slightly better than HPRA(D). 
Both SOPPA and HPRA(D) mostly overestimate the values, which is notable in the mean deviation. 
The consistency (std dev) and the accuracy (mean abs dev) for the two methods are close, but SOPPA performs also slightly better for these.  
The SSCCs with the largest deviation are $^1J_{CF}(\text{F}_2\text{CO})$ for SOPPA and $^1J_{CF}(\text{FCCF})$ for HRPA(D). 

By looking at the individual SSCCs, it notable that 55 out of 72 SSCCs deviate less than 10 Hz from the experimental values for both SOPPA and HRPA(D).
When the deviation limit is set to 5 Hz, 44 out of 72 SSCCs fall within this limit for SOPPA and 47 SSCCs of out 72 are within for HRPA(D).

The data set contains the three simplest hydrocarbons with a C-C single, double and triple bond. We can therefore compare how well the methods perform when the bond changes from a single bond to a triple bond. The deviations from experimental values,  ($J_{calc}-J_{exp}$), calculated for ethane, ethene, and ethyne are presented in Table \ref{table:deviation_ethane-ethene-ethyne}.

\begin{table}[h!]
    \caption{Deviation from experimental values (in Hz) for the SSCCs of ethane, ethene, and ethyne.}
    \label{table:deviation_ethane-ethene-ethyne}
\centering
\setlength{\tabcolsep}{15pt}
\begin{tabular}{lcrr}
\hline
 \textbf{SSCC} & \textbf{Molecule} & \textbf{SOPPA} & \textbf{HRPA(D)}\\ \hline
$^1J_{CC}$ & Ethyne & 20.18 & 17.98  \\
 & Ethene & 3.48 & 3.60  \\ 
 & Ethane & 1.69 & 2.33  \\ \hline
$^1J_{CH}$ & Ethyne & 15.67 & 14.38  \\
 & Ethene & 4.63 & 4.16  \\
 & Ethane & 1.18 & 1.22  \\ \hline
$^3J_{HH}$ & Ethyne & 2.77 & 2.05  \\
\hspace{0.2cm} \footnotesize{(trans)} & Ethene  & 0.07 & -0.82  \\
\hspace{0.2cm} \footnotesize{(cis)} & Ethene  & 1.12 & 0.51  \\ \hline
$^2J_{CH}$ & Ethyne & 2.54 & 1.33  \\ \hline
\end{tabular}
\end{table}

\begin{table*}[hbpt]
    \caption{Statistical analyses of the deviations from the experimental values (in Hz) of the molecules in set II, divided into groups of bond types.}
    \label{table:Statistics_set2-groups-of-bonds}
\centering
\begin{tabular}{l|rr|rr|rr}
\hline
 & \multicolumn{2}{c|}{\textbf{Single-bond}} & \multicolumn{2}{c|}{\textbf{Double-bond}} & \multicolumn{2}{c}{\textbf{Triple-bond}} \\
 & \textbf{SOPPA} & \textbf{HRPA(D)} & \textbf{SOPPA} & \textbf{HRPA(D)} & \textbf{SOPPA} & \textbf{HRPA(D)} \\ \hline
Count & \multicolumn{1}{c}{15} & \multicolumn{1}{c|}{15} & \multicolumn{1}{c}{33} & \multicolumn{1}{c|}{33} & \multicolumn{1}{c}{24} & \multicolumn{1}{c}{24} \\
Mean Abs Dev & 3.73 & 3.27 & 6.85 & 8.01 & 7.08 & 6.29 \\
Mean Dev & 1.36 & 2.49 & -0.24 & 3.61 & 3.30 & 5.22 \\
Std Dev & 5.66 & 4.60 & 9.56 & 11.25 & 8.97 & 7.98 \\
Max Abs Dev & 15.62 & 17.58 & 22.28 & 29.55 & 21.72 & 29.63 \\ \hline
\end{tabular}
\end{table*}

For these simple hydrocarbons, it is generally found that the deviations increase when going from a single bond to a triple bond for both methods. 
It is notable that SOPPA performs best for the coupling constants of ethane, HPRA(D) performs best for the coupling constants of ethyne, and for the coupling constants of ethene, they alternate to perform best. 

It would be interesting to see, if this holds in general or whether this applies only for the simple hydrocarbons. 
Therefore in the following, all 31 molecules will be investigated. 
All the molecules have been divided into groups according to which bonds the molecule contains with the highest bond order counting. 
The statistical analyses is presented in Table \ref{table:Statistics_set2-groups-of-bonds}.

For the 10 molecules only containing single bonds, HPRA(D) performs best.
Both methods mostly overestimate the values of the SSCCs.
In accuracy, the two methods are quite close with HRPA(D) being slightly better.
In consistency, HRPA(D) performs considerably better than SOPPA. 
The coupling constants with the largest deviation from experimental values are $^1J_{NF}$(F$_2$NNF$_2$) for SOPPA and $^1J_{CF}$(CF$_4$) for HRPA(D). 
12 out of 15 SSCCs are within 3 Hz from the experimental values for both SOPPA and HRPA(D). The outliers are the coupling constants $^1J_{HH}$(H$_2$) and $^1J_{CF}$(CF$_4$) for both methods, $^1J_{NF}$(F$_2$NNF$_2$) for only SOPPA, and $^1J_{NH}$(NH(CH$_3$)$_2$) for only HRPA(D) with the last one being within 5 Hz from experimental value.
When comparing SOPPA and HRPA(D) for the individual SSCCs, the two methods perform closely to each other with 13 out of 15 SSCCs differing less than 1 Hz. The 2 SSCCs that differ more are $^1J_{NF}$(F$_2$NNF$_2$) with a difference at 14.59 Hz and $^1J_{CF}$(CF$_4$) with a difference at 30.89 Hz. 

SOPPA performs best for the 33 coupling constants of the 13 molecules containing at least one double bond. 
SOPPA mostly underestimates the values whereas HRPA(D) mostly overestimates them.  
The SSCCs of SOPPA are more consistent and accurate than those of HRPA(D). 
The coupling constants with the largest deviation from experimental values are $^1J_{CF}(\text{F}_2\text{CO})$ for SOPPA and $^3J_{FF}(\text{trans-FHCCHF})$ for HRPA(D).

For SOPPA, 30 out of 33 SSCCs are within 20 Hz from the experimental values, and 26 out of 33 SSCCs are within for HRPA(D). 
The 3 SSCCs that deviate more than 20 Hz for SOPPA are $^1J_{CF}(\text{F}_2\text{CO})$ (deviates -22.28 Hz), $^1J_{NF}$(FNNF) for both cis and trans (deviate 22.08 Hz and 20.58 Hz, respectively). 
For HRPA(D), 7 SSCCs deviate more than 20 Hz from experimental values, $^1J_{CF}$ for F$_2$CO and F$_2$CCH$_2$ (deviation of 20.04 Hz and 20.96 Hz, respectively), $^3J_{FF}$ for F$_2$CCF$_2$ for both cis and trans coupling, and trans-FHCCHF (deviation of 21.57 Hz, 24.81 Hz, and 29.55 Hz, respectively), $^2J_{NF}$(cis-FNNF) (deviation of 20.33 Hz), and $^2J_{FF}$(F$_2$CNH) (deviates 22.13 Hz). 
5 SSCCs for SOPPA deviate between 10 and 20 Hz from experimental values: $^2J_{FF}$(F$_2$CCF$_2$) (deviates -16.83 Hz), $^1J_{CF}$(F$_2$CCH$_2$) (deviates -14.28 Hz), $^2J_{FF}$(F$_2$CCH$_2$) (deviates -18.42 Hz), $^2J_{NF}$(cis-FNNF) (deviates 12.03 Hz) and $^2J_{FF}$(F$_2$CNH) (deviates -14.69 Hz). 
For HRPA(D), 3 SSCCs fall within this range: $^2J_{FF}$(F$_2$CCF$_2$) (deviates -18.03 Hz), $^2J_{NF}$(trans-FNNF) (deviates 13.17 Hz), and $^1J_{NF}$(cis-FNNF)(deviates -11.80 Hz). 
For the individual SSCCs, SOPPA and HRPA(D) perform similarly with 22 out of 33 SSCCs being within 4 Hz of each other. The coupling constants with a larger difference are 
$^1J_{CF}(\text{F}_2\text{CO})$ (differ 42.32 Hz), 
$^3J_{FF}$(F$_2$CCF$_2$) (differ 28.14 Hz and 29.99 Hz for trans and cis, respectively), 
$^1J_{CF}$(F$_2$CCH$_2$)(differ 35.24 Hz), 
$^2J_{FF}$(F$_2$CCH$_2$) (differ 26.25 Hz), 
$^3J_{FF}$(trans-FHCCHF) (differ 36.34 Hz), 
$^1J_{NF}$(FNNF) (differ 23.20 Hz and 33.87 Hz for trans and cis, respectively), 
$^2J_{NF}$(FNNF) (differ 9.29 Hz and 8.30 Hz for trans and cis, respectively), 
and $^2J_{FF}$(F$_2$CNH) (differ 36.82 Hz).

For the 24 coupling constant of the 8 molecules containing a triple bond, HPRA(D) performs best. 
HPRA(D) overestimates almost all the values, whereas SOPPA underestimates some of the values more.
Both in consistency and accuracy HRPA(D) performs better than SOPPA. 
For SOPPA the coupling constant with the largest deviation from experimental values is $^3J_{FF}(\text{FCCF})$ and for HRPA(D) it is $^1J_{CF}(\text{FCCF})$. 
When checking the individual SSCCs, it is found that within a 10 Hz range from experimental values, SOPPA has 17 out of 24 SSCCs and HRPA(D) has 18 out of 24 SSCCs. The coupling constants which have a larger deviation are $^1J_{CH}(\text{HCCH})$, $^1J_{CC}(\text{HCCH})$, $^1J_{CF}(\text{FCCF})$, $^1J_{CC(\equiv)}$(H$_3$CCCH) and $^1J_{CC(CH)}$(H$_3$CCCH) for both methods, 
$^3J_{FF}(\text{FCCF})$ and $^2J_{CF}(\text{FCCF})$ for only SOPPA, and $^3J_{FH}(\text{FCCH})$ for only HRPA(D). 
For 22 out of 24 SSCCs, SOPPA and HRPA(D) perform within 4 Hz of each other.
The last two SSCCs are $^3J_{FF}(\text{FCCF})$ and $^1J_{CF}(\text{FCCF})$ with a difference of 23.66 Hz and 41.77 Hz, respectively.

\subsection{Set III: Aromatic and fluoro-aromatic compounds}
The 368 different spin-spin coupling constants of the 27 aromatic and fluoro-aromatic compounds have been calculated with the methods SOPPA and HRPA(D). All the SSCCs including the experimental values 
\cite{book_Berger1997_non-metallic-elements-NMR_exp-values, book_kalinowski1988_carbon13-NMR-spec_exp-values, book_stothers1972_carbon13-NMR-spec_exp-values, Roznyatovsky1991_exp-value-benzene, KAMIENSKATRELA1995_Benzene-difluorobenzene, Tori1964_1J_CH, Wray1977_fluoro-sub-benzene_FH-FF-CH-CF-couplings, JOHNSON1996_exp-value_pyridine, PALMER1997_exp-value_1245-tetrazine, Ernst1976_12-difluorobenzene_7c-8e, Ernst1977_fluorobenzene, KKAMIENSKATRELA1993_12-difluorobenzene-CC-couplings, Abraham1967_123-trifluorobenzene, Abraham1968_123-trifluorobenzene, WRAY1975_135-TRIFLUOROBENZENE, Darabantu2001_36-difluoropyridazine, Bobbio2005_345-Trifluoropyridine, Gross2000_246-Trifluoro-135-triazine}
can be found in Table S4 in supplementary materials.

\begin{figure}[h!]
    \centering
    \includegraphics[width=0.5\textwidth]{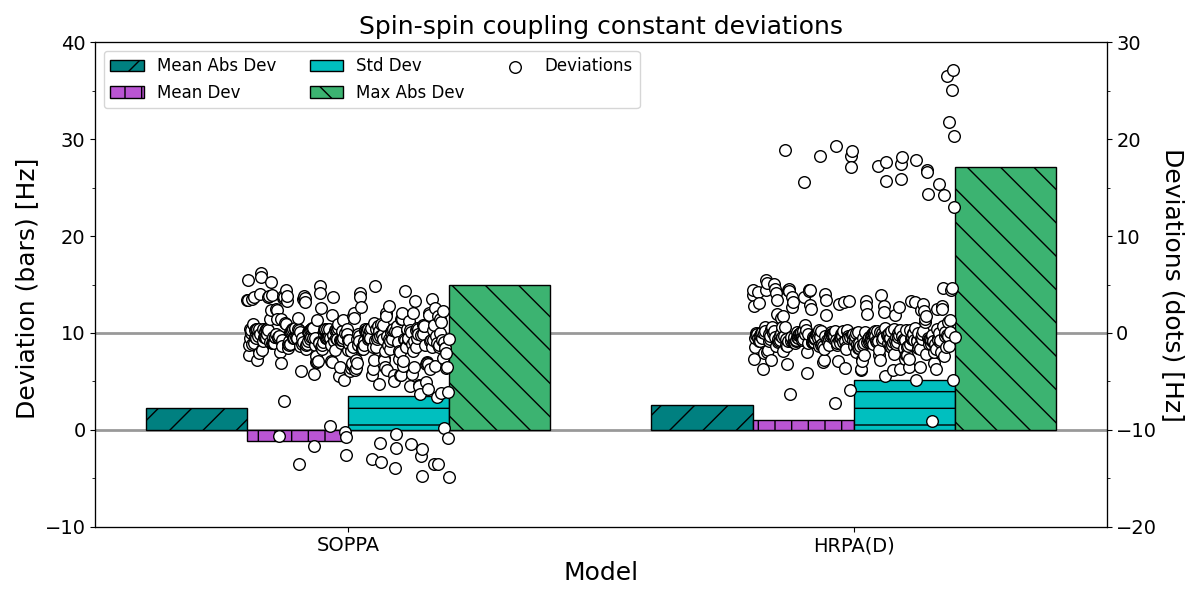}
    \caption{Statistical analysis of the deviations from the experimental values (in Hz) for all methods and all molecules in set III. The dots indicated the deviations of the individual couplings.}
    \label{fig:MP2-geo_all-aromatic}
\end{figure}

A statistical analysis has been performed with the experimental values as the reference. The results are presented in Figure \ref{fig:MP2-geo_all-aromatic} and Table \ref{tab:MP2-aromatic-all}. 
The one-bond coupling constants between carbon and fluorine ($^1J_{CF}$) are resulting in outliers, where SOPPA underestimates and HRPA(D) overestimates the values. 
The 25 $^1J_{CF}$ coupling constants have a mean deviation at -11.29 Hz for SOPPA with deviations between -14.90 and -3.58, and for HRPA(D) the mean deviation is 18.30 Hz with deviations between 12.98 and 27.16. 
These coupling constants affect the statistical analysis and therefore a second analysis has been performed without them which can be found in Table \ref{tab:MP2-aromatic-all} as well.

\begin{table}[h!]
    \caption{Statistical analysis of the deviations from the experimental values (in Hz) for all methods and all molecules in set III with and without the $^1J_{CF}$ coupling constants.}
    \label{tab:MP2-aromatic-all}
\centering
\begin{tabular}{l|rr|rr}
\hline
 & \multicolumn{2}{c|}{\textbf{All SSCCs}} & \multicolumn{2}{c}{\textbf{Without $^1J_{CF}$}} \\ 
  & \textbf{SOPPA} & \textbf{HRPA(D)} & \textbf{SOPPA} & \textbf{HRPA(D)} \\ \hline
Count & 368 & 368  & 343 & 343 \\
Mean Abs Dev & 2.28 & 2.54 & 1.62 & 1.39 \\
Mean Dev & -1.10 & 1.01 & -0.36 & -0.25\\
Std Dev & 3.54 & 5.13 & 2.19 & 1.99 \\
Max Abs Dev & 14.90 & 27.16 & 7.01 & 9.10\\ \hline
\end{tabular}
\end{table}

The two methods perform quite closely to each other both with and without the $^1J_{CF}$ coupling constants. 
When looking at all the SSCCs, SOPPA is performing slightly better than HRPA(D). 
Both methods have low mean (abs.) Dev. with SOPPA having a slightly lower one. 
SOPPA has a better consistency than HRPA(D), i.e. smaller Std. Dev., which is not surprising since HRPA(D) overestimates the $^1J_{CF}$ coupling constants more than SOPPA underestimates them. 
The SSCCs that deviate the most from experimental values are the $^1J_{CF}$ from the molecules 2,4,6-trifluoro-1,3,5-triazine and 3,4,5-trifluoropyridine for SOPPA and HRPA(D), respectively. 
When the $^1J_{CF}$ coupling constants are ignored, both methods have an even smaller mean (abs) dev and Std. Dev. with HRPA(D) performing slightly better. 

Looking at the individual SSCCs, most of the SSCCs are within 5 Hz from the experimental values. For SOPPA, 331 out of 368 SSCCs are within 5 Hz for which 24 of the outliers are the $^1J_{CF}$ coupling constants. For HRPA(D), 336 out of 368 SSCCs are within 5 Hz for which 25 of the outliers are the $^1J_{CF}$ coupling constants.

In the next sections, the 27 molecules have been divided into three groups: Aromatic (molecules without any fluorine), fluoro-aromatic (molecules with fluorine but no nitrogen), and fluoro-nitro-aromatic (molecules with both fluorine and nitrogen). 
For every group, a statistical analysis was performed with the experimental values used as references to see which type of method performs best for each type of molecule.

\subsubsection{Aromatic}
\begin{figure}[h!]
    \centering
    \includegraphics[width=0.5\textwidth]{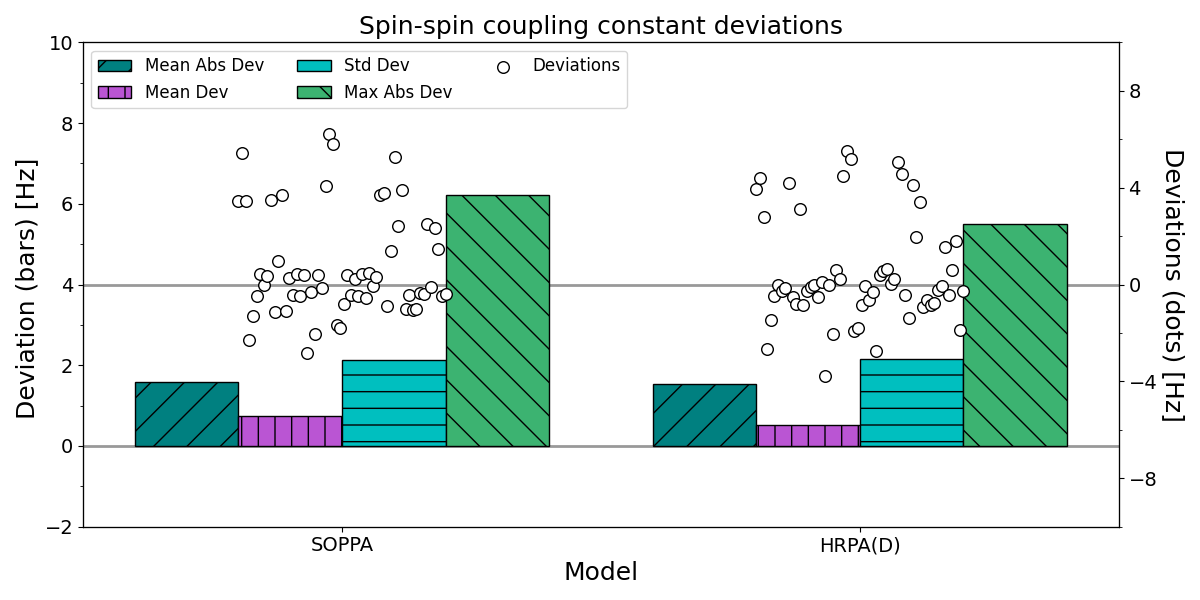}
    \caption{Statistical analysis of the deviations from the experimental values (in Hz) for all methods and all molecules that do not contain fluorine in set III. The dots indicated the deviations of the individual couplings.}
    \label{fig:MP2-geo_aromatic-aromatic}
\end{figure}

For the 7 aromatic molecules, furan, benzene, pyrrole, pyridine, pyrazine, 1,3,5-triazine, and 1,2,4,5-tetrazine, the statistical analysis is presented in Figure \ref{fig:MP2-geo_aromatic-aromatic} and Table \ref{tab:MP2-aromatic-aromatic}. 

\begin{table}[h!]
    \caption{Statistical analysis of the deviations from the experimental values (in Hz) for all methods and all molecules that do not contain fluorine in set III.}
    \label{tab:MP2-aromatic-aromatic}
\centering
\begin{tabular}{lcc}
\hline
 & \textbf{SOPPA} & \textbf{HRPA(D)} \\ \hline
Count & 58 & 58 \\
Mean Abs Dev & 1.58 & 1.54 \\
Mean Dev & 0.74 & 0.51 \\
Std Dev & 2.14 & 2.17 \\
Max Abs Dev & 6.21 & 5.50 \\ \hline
\end{tabular}
\end{table}

For the aromatic molecules, the two methods perform similarly well. 
HRPA(D) performs slightly better in accuracy and SOPPA performs slightly better in consistency with both being within 0.04 of each other, and therefore one would not be considered better than the other. However, since HRPA(D) is less expensive than SOPPA, it would be the preferred choice. 
The largest deviation from experimental values originates from the same SSCC, the $^1J_{CH}$ from the molecule pyrrole.  

For SOPPA, 54 out of 58 SSCCs are within 5 Hz from experimental values. The four which deviate more than 5 Hz, are all $^1J_{CH}$ coupling constants from the molecules furan (C2: 5.43 Hz), pyrrole (C2: 6.21 Hz, C3: 5.82 Hz) and pyridine (C2: 5.28 Hz).
For HRPA(D), 55 out of 58 SSCCs lie within 5 Hz from experimental values. The 3 outliers are two $^1J_{CH}$ coupling constants from pyrrole with deviations 5.50 Hz (C2) and 5.20 Hz (C3), and $^1J_{C2C3}$ coupling constant from pyridine with a deviation of 5.05 Hz. 

By looking at the individual SSCCs, SOPPA and HRPA(D) perform closely to each other with a difference not extending a few Hz (the largest difference is 2.77 Hz for $^3J_{CC}$(pyridine)) and mostly within one Hz from each other (only 9 of the 58 SSCCs have a larger difference).

\subsubsection{Fluoro-aromatic}
\begin{figure}[h!]
    \centering
    \includegraphics[width=0.5\textwidth]{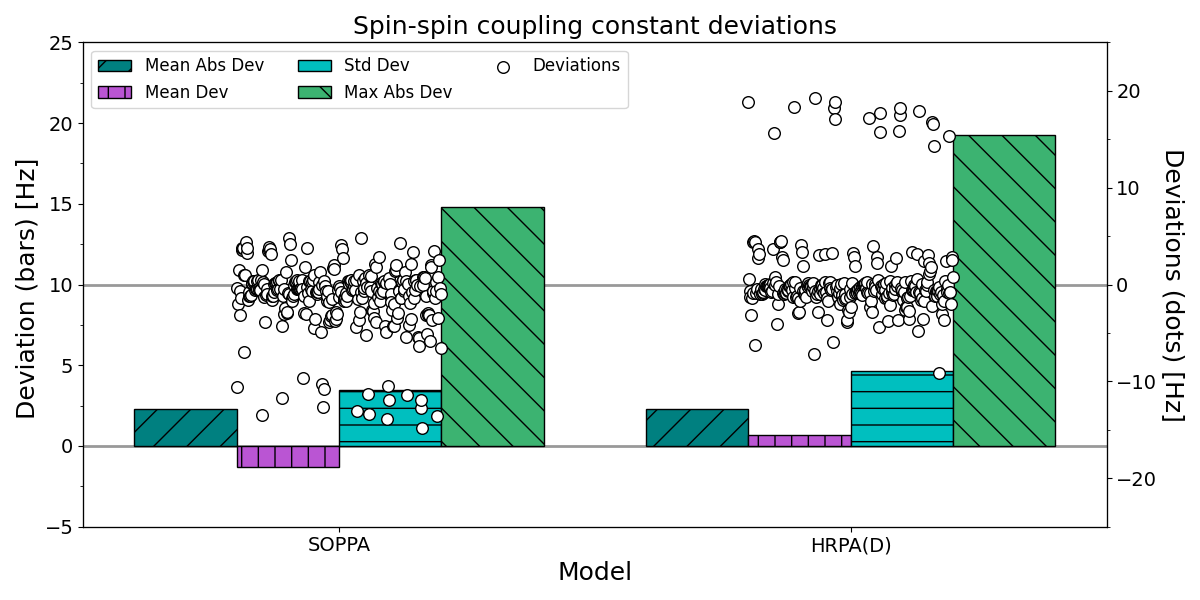}
    \caption{Statistical analysis of the deviations from the experimental values (in Hz) for all methods and all molecules that do contain fluorine but no nitrogen in set III. The dots indicated the deviations of the individual couplings.}
    \label{fig:MP2-geo_fluoro-aromatic-aromatic}
\end{figure}

For the 12 fluoro-aromatic molecules: fluorobenzene, 1,2-difluorobenzene, 1,3-difluorobenzene, 1,4-difluorobenzene, 1,2,3-trifluorobenzene, 1,2,4-trifluorobenzene, 1,3,5-trifluorobenzene, 1,2,3,4-tetrafluorobenzene, 1,2,3,5-tetrafluorobenzene, 1,2,4,5-tetra-fluorobenzene, 1,2,3,4,5-pentafluorobenzene, and hexafluorobenzene, the statistical analysis is presented in Figure \ref{fig:MP2-geo_fluoro-aromatic-aromatic} and Table \ref{tab:MP2-aromatic-fluoro-aromatic}. 

\begin{table}[h!]
    \caption{Statistical analysis of the deviations from the experimental values (in Hz) for all methods and all molecules that do contain fluorine but no nitrogen in set III with and without the $^1J_{CF}$ coupling constants.}
    \label{tab:MP2-aromatic-fluoro-aromatic}
\centering
\begin{tabular}{l|rr|rr}
\hline
 & \multicolumn{2}{|c|}{\textbf{All SSCCs}} & \multicolumn{2}{c}{\textbf{Without $^1J_{CF}$}} \\ 
  & \textbf{SOPPA} & \textbf{HRPA(D)} & \textbf{SOPPA} & \textbf{HRPA(D)} \\ \hline
Count & 288 & 288 & 270 & 270 \\
Mean Abs Dev & 2.29 & 2.32 & 1.63 & 1.33 \\
Mean Dev & -1.30 & 0.66 & -0.58 & -0.44\\
Std Dev & 3.49 & 4.65 & 2.15 & 1.86 \\
Max Abs Dev & 14.81 & 19.26 & 7.01 & 9.10 \\ \hline
\end{tabular}
\end{table}

The parameters of the statistical analysis are close to each other for SOPPA and HRPA(D) both with and without $^1J_{CF}$ coupling constants.
For all the SSCCs, SOPPA performs slightly better with lower values in both mean (abs.) Dev. and Std. Dev. 
However, when excluding the $^1J_{CF}$ coupling constants, HRPA(D) performs slightly better.
Both methods have a tendency to mostly underestimate the values compared with the experimental values when excluding the $^1J_{CF}$ coupling constants.
The largest deviations are different SSCCs for SOPPA and HRPA(D). 
For all the SSCCs, the largest deviations are from the $^1J_{CF}$ coupling constants of the molecules 1,2,3,4,5-pentafluorobenzene and 1,4-difluorobenzene for SOPPA and HRPA(D), respectively. 
Without the $^1J_{CF}$ coupling constants, the largest deviation from experimental values are the coupling constants $^2J_{CC}$(fluorobenzene) for SOPPA and $^3J_{CF}$(1,2,3,4,5-pentafluorobenzene) for HRPA(D). 

When looking at a deviation for experimental values at 5 Hz, SOPPA has 261 out of 288 SSCCs and HRPA(D) has 266 out of 288 SSCCs within the range. 18 among the outliers are $^1J_{CF}$ coupling constants for both methods meaning that SOPPA has 9 other outliers and HRPA(D) has 4 other outliers underestimating the values. 
The 9 outliers for SOPPA are of four different types, the 
$^2J_{C2C4}$ from the molecule fluorobenzene (deviates -7.01 Hz), $^4J_{FF}$ from the molecules 1,2,3,4-tetrafluorobenzene (F1F3: \mbox{-5.24 Hz}), 
1,2,4,5-tetrafluorobenzene (-5.44 Hz), 
1,2,3,4,5-pentafluorobenzene (F1F3: -5.37 Hz , F1F5: -5.52 Hz, F2F4: -6.30 Hz), and 
hexafluorobenzene (-6.58 Hz),  
$^3J_{C1F5}$ and $^4J_{C1F4}$ from the molecule 1,2,3,4,5-pentafluorobenzene (-5.06 Hz and -5.80 Hz, respectively). 
For HRPA(D) the outliers are $^2J_{C2C4}$ from fluorobenzene (deviates -6.28 Hz), $^5J_{FF}$ from 1,4-difluorobenzene and 1,2,4-trifluorobenzene (deviate -7.22 Hz and -5.88 Hz, respectively), and $^3J_{C1F5}$ from 1,2,3,4,5-pentafluorobenzene (deviates -9.10 Hz). 

The 18 $^1J_{CF}$ coupling constants for the fluoro-aromatics have a mean deviation from experimental values for SOPPA of -12.08 Hz with individual deviations between -14.81 Hz and \mbox{-9.61} Hz and for HRPA(D) the mean deviation of the $^1J_{CF}$ coupling constants is 17.19 Hz for a range of values from 14.35 Hz to 19.26 Hz. 
Earlier the mean deviations for all 25 $^1J_{CF}$ coupling constants were calculated, comparing them to the 18 $^1J_{CF}$ coupling constants in this set of molecules, the values are lower for both SOPPA and HRPA(D) which means that for the fluoro-aromatic, SOPPA underestimates the $^1J_{CF}$ SSCCs more and HRPA(D) overestimated the $^1J_{CF}$ SSCCs less.  

For the individual SSCCs, SOPPA and HRPA(D) perform closely to each other except for 18 $^1J_{CF}$ coupling constants, where the mean difference is 29.27 Hz with differences between 27.93 Hz and 30.24 Hz. 
When excluding the $^1J_{CF}$ coupling constants,
only 10 SSCCs for SOPPA and HRPA(D) differ by more than 5 Hz from each other, which are some of the long-range couplings between two fluorine atoms ($^4J_{FF}$ and $^5J_{FF}$). The largest difference is 9.06 Hz for $^4J_{FF}$ from hexafluorobenzene. 
58 out of 270 SSCCs have a larger than 2 Hz difference between the SOPPA and HRPA(D) calculation (which includes the 10 SSCCs above a difference at 5 Hz).

\subsubsection{Fluoro-nitro-aromatic}
\begin{figure}[h!]
    \centering
    \includegraphics[width=0.5\textwidth]{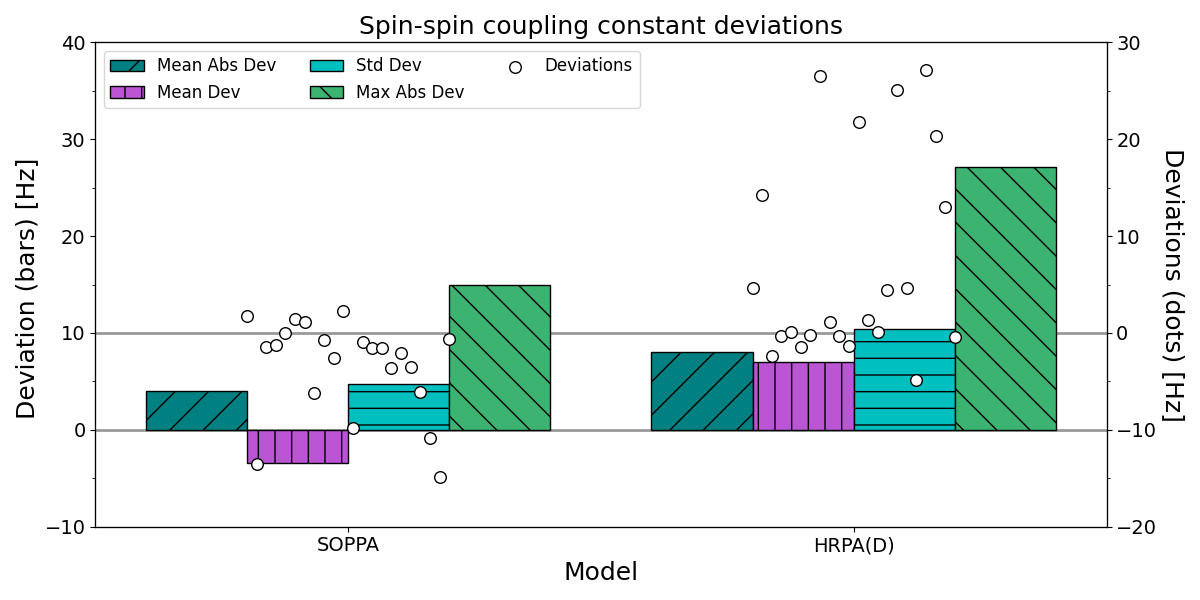}
    \caption{Statistical analysis of the deviations from the experimental values (in Hz) for all methods and all molecules that do contain both fluorine and nitrogen in set III. The dots indicated the deviations of the individual couplings.}
    \label{fig:MP2-geo_fluoro-nitro-aromatic-aromatic}
\end{figure}

For the 8 fluoro-nitro-aromatic molecules: 2-fluoropyri-dine, 3-fluoropyridine, 4-fluoropyridine, 2,6-difluoropyridine, 3,6-difluoropyridazine, 2,4,6-trifluoropyridine, 3,4,5-trifluoro-pyridine, and 2,4,6-trifluoro-1,3,5-triazine, the statistical analysis is presented in Figure \ref{fig:MP2-geo_fluoro-nitro-aromatic-aromatic} and Table \ref{tab:MP2-aromatic-fluoro-nitro-aromatic}.

\begin{table}[h!]
    \caption{Statistical analysis of the deviations from the experimental values (in Hz) for all methods and all molecules that do contain both fluorine and nitrogen in set III with and without the $^1J_{CF}$ coupling constants.}
    \label{tab:MP2-aromatic-fluoro-nitro-aromatic}
\centering
\begin{tabular}{l|rr|rr}
\hline
 & \multicolumn{2}{|c|}{\textbf{All SSCCs}} & \multicolumn{2}{c}{\textbf{Without $^1J_{CF}$}} \\ 
  & \textbf{SOPPA} & \textbf{HRPA(D)} & \textbf{SOPPA} & \textbf{HRPA(D)} \\ \hline
Count & \multicolumn{1}{c}{22} & \multicolumn{1}{c|}{22} & \multicolumn{1}{c}{15} & \multicolumn{1}{c}{15} \\
Mean Abs Dev & 3.99 & 7.99 & 1.53 & 1.85 \\
Mean Dev & -3.40 & 6.97 & -0.66 & 0.34 \\
Std Dev & 4.75 & 10.37 & 1.62 & 2.55 \\
Max Abs Dev & 14.90 & 27.16 & 3.55 & 4.82\\ \hline
\end{tabular}
\end{table}

When examining all the SSCCs, SOPPA performs significantly better than HRPA(D) which is notable in the accuracy (mean (abs) dev) and the consistency (std dev) where the values for SOPPA are almost half of the values of HRPA(D). 
SOPPA generally underestimates the SSCCs, while HRPA(D) generally overestimates them.
The largest deviations from experimental values are the same as for all 368 SSCCs, the $^1J_{CF}$ coupling constants from the molecules 2,4,6-trifluoro-1,3,5-triazine and 3,4,5-trifluoropyridine for SOPPA and HRPA(D), respectively.

When not considering the $^1J_{CF}$ coupling constants, SOPPA performs slightly better than HRPA(D) in accuracy, whereas for consistency SOPPA performs noticeably better than HRPA(D), however, it should be noted that the values both are in the lower end. 
The largest deviation from experimental values is the same coupling constant for both methods, the $^4J_{F2F4}$ from the molecule 2,4,6-trifluoropyridine. 

Looking at the individual SSCCs, SOPPA has 14 out of 22 SSCCs within 3 Hz deviation from experimental values, whereas HRPA(D) has 11 out of 22 SSCCs. 
Of these outliers, 7 arise from the $^1J_{CF}$ coupling constants, which means when excluding the $^1J_{CF}$ coupling constants, SOPPA has one SSCC and HRPA(D) has four SSCCs that deviate more than 3 Hz. 
The outlier for SOPPA is the $^4J_{F2F4}$ coupling constant from 2,4,6-trifluoropyridine (deviates -3.55 Hz). 
The four outliers for HRPA(D) are the $^2J_{NF}$ coupling constant from 2-fluoropyridine (deviates 4.67 Hz), the three $^4J_{FF}$ from the molecules 2,6-difluoropyridine (4.47 Hz) and 2,4,6-trifluoropyridine (F2F6: 4.63 Hz, F2F4: -4.82 Hz).

The mean deviation from experimental values for the 7 $^1J_{CF}$ coupling constants for the fluoro-nitro-aromatics is for SOPPA -9.27 Hz with values between -14.90 Hz and -3.58 Hz and for HRPA(D) it is 21.17 Hz with values between 12.98 Hz and 27.16 Hz. 
When comparing the mean deviation for the 7 $^1J_{CF}$ coupling constants for the fluoro-nitro-aromatics to the 25 $^1J_{CF}$ coupling constants, the values for SOPPA and HRPA(D) increase, meaning that SOPPA underestimates the values less and HRPA(D) overestimates the values more.

For the individual SSCCs the difference between SOPPA and HRPA(D) are not large. 
For the 7 $^1J_{CF}$ coupling constants the mean difference is 30.43 with differences between 27.71 and 33.27.
Without the $^1J_{CF}$ coupling constants, 
for 7 out of 15 SSCCs the SOPPA and HRPA(D) values differ by more than 2 Hz from each other, and for 2 SSCCs by more than 5 Hz: $^4J_{FF}$ in 2,6-difluoropyridine and 2,4,6-trifluoropyridine.

For the fluoro-nitro-aromatics, SOPPA would be the preferred method if the calculation includes the $^1J_{CF}$ coupling constant, whereas if the $^1J_{CF}$ coupling constant is not considered, both SOPPA and HRPA(D) would be adequate methods with SOPPA performing slightly better and HRPA(D) being a cheaper method.

\section{Conclusion}
This study aimed to investigate how well the method HRPA(D) performs for spin-spin coupling constants for smaller molecules, aromatic and fluoro-aromatic compounds compared with SOPPA in comparison to experimental data.
Typical, the two methods perform closely to each other and alternate to perform best. 

SSCC calculations were performed on three sets of molecules for both SOPPA and HRPA(D). 
For set I (7 smaller molecules), two different values were used as references in the statistical analysis: experimental values and calculated CC3 values.
SOPPA was closest to the experimental values and even performed better than CC3 and CCSD. HRPA(D) performed closely to CC3 and CCSD. 
CCSD was closest to the CC3 values. Of SOPPA and HRPA(D), HRPA(D) performed better.

For set II (31 smaller molecules), SOPPA and HRPA(D) perform closely with SOPPA performing slightly better when compared to the experimental values. 
When dividing the set, it was found that for molecules only containing single bonds and molecules having one triple bond, HRPA(D) performs slightly better than SOPPA. However, for molecules containing double bond(s), SOPPA performs better than HRPA(D).

For set III (27 aromatic and fluoro-aromatic compounds), one type of coupling constant accounted for the largest outliers, the $^1J_{CF}$ coupling constants. For this coupling constant, SOPPA underestimated less than HRPA(D) overestimated the value compared to the experimental values.
When excluding the $^1J_{CF}$ coupling constants, HRPA(D) overall performs slightly better than SOPPA. 
Dividing this set into aromatics, fluoro-aromatics, and fluoro-nitro-aromatics, the methods alternate in performing best. For the aromatics, both methods perform similarly well. For the fluoro-aromatics, HRPA(D) performs slightly better than SOPPA. For the fluoro-nitro-aromatics, SOPPA performs better than HRPA(D). 

HRPA(D) is an approximation to SOPPA, however, it performs comparable and sometimes better, and with it being a computationally less expensive method, it should be considered when calculating spin-spin coupling constants.

In this study, it was further investigated which geometry optimization (CCSD(T) vs MP2) resulted in the best SSCCs compared with experimental values.
For SOPPA, the more precise calculated SSCCs were obtained with a CCSD(T) geometry optimization.  
For HRPA(D), the MP2 geometry optimization resulted in spin-spin coupling constants that are closer to the experimental values.

%

\end{document}